\documentstyle[12pt]{article}
 \advance
 \oddsidemargin by
 -1in
 \advance
 \evensidemargin
 by -1in
 \advance
 \topmargin by -0.0in
 \makeindex
 \newcommand
 \beq{\begin{equation}}
 \newcommand\eeq{\end{equation}}
 \newcommand\beqn{\begin{eqnarray}}
 \newcommand\eeqn{\end{eqnarray}}
 \newcommand{\doublespace} {
 \renewcommand{\baselinestretch} {1.6}
 \large\normalsize}
 
\begin{document}
%\vspace*{0.5cm}
%\hspace*{9cm}{\Large\bf MPIH-V12-1995}
\vspace*{2cm}
 
 \centerline{{\huge\bf Scaling Variable }}
\medskip
 \centerline{{\huge\bf for Nuclear Shadowing}}
\medskip
 \centerline{\huge {\bf in Deep-Inelastic Scattering}\footnote
{Based on the  talk
 presented by B.~Kopeliovich at the Workshop
on Particle Theory and Phenomenology, Ames, Iowa, 1995}}
\vspace{1.5cm}
\begin{center}
{\large Boris~Kopeliovich\footnote{On leave from
 the Joint Institute for
 Nuclear Research,
 Laboratory of Nuclear Problems, \newline Dubna,
 141980
 Moscow Region, Russia. E-mail:
 bzk@dxnhd1.mpi-hd.mpg.de}$\ $ and
 Bogdan~Povh}
 \date{}
 
\vspace{0.5cm}
 
 {\sl Max-Planck-Institut f\"ur Kernphysik,
 Postfach
 103980, \newline 69029 Heidelberg,
 Germany}\\ \end{center} \bigskip
 
\doublespace
\begin{abstract}

 A new scaling variable is introduced in terms of
 which nuclear shadowing
 in deep- inelastic
 scattering is universal, i.e. independent of $A$,
 $Q^2$
 and $x$. This variable can be interpreted
 as a measure of the
 number
 of
 gluons probed by the
 hadronic fluctuations of a virtual photon
 during their lifetime.  
The shadowing correction grows at small $x$
 substantially less steeply than is suggested by
the eikonal approximation. This results from 
the fact
 that shadowing is dominated by soft
 rather than hard interactions.

\end{abstract}
\newpage
 
 1. {\bf Introduction} \\
 A large amount of new high precision
 data on
 nuclear
 shadowing in deep-inelastic
 scattering is
 now
 available.  The
 comparison with
 theory is not easy since the
 data are
 taken
 for a variety
 of nuclei at different values of $Q^2$
 and
 $x$. One of the goals of 
 this paper is
 to find a scaling variable,
 dependent
 on $Q^2$, $x$
 and $A$, which makes
 shadowing an universal function of
 this
 variable.
 Our guess of the correlations between $Q^2$, $x$
 and $A$
 entering
 the
 scaling variable is inspired
 by the prejudice about the
 underlying QCD
 dynamics
 of shadowing.
 
 Deep-inelastic scattering is
 usually
 interpreted
 in two alternative frames of reference, in the
 infinite
 momentum
 frame of the proton in
 terms of its structure
 function, or, in
 the proton rest frame in terms of
 hadronic fluctuations
 of the photon.  In
 the first case, nuclear
 shadowing looks like a result of
 overlap in the
 longitudinal
 direction of the parton clouds originated from
 different bound
 nucleons.
 The latter approach is more in
 line with the
 familiar
 vector
 dominance model \cite{bauer}.  
It seems to
 be advantageous treating
 shadowing at small $x$
as the consequence of 
the hadronic fluctuations of the photon.
 On the technical level
 our
 approach is similar to
 that of \cite{nz91}.
 
Prior to the detailed
 discussion,
 we introduce the
 scaling variable 
motivated by the Glauber model \cite{glauber}
as an average
 number of bound nucleons taking part in 
interacting with the
 hadronic
 fluctuation
 of the
 photon,
 
 \beq
 n(x,Q^2,A)={1\over 4}
 \frac{\langle\sigma_h^2\rangle}
 {\langle\sigma_h \rangle}
 \langle
 T_A\rangle F^2_A(q_L)
\label{1}
\eeq
 
 In this expression $\sigma_h$ is the total cross
 section of
 interaction of the hadronic fluctuation
 $h$ of the photon with a nucleon.
 We
 choose the
 basis of states $h$ to be eigenstates of
the interaction,
 rather
 than
 of the mass matrix.  This
 eliminates the off diagonal
 amplitudes in
 the
 double dispersion relation \cite{gribov1} for
 deep-inelastic
 scattering.  The concrete choice of
the eigenstate basis is done below.
The
 averaging in eq.
 (\ref{1}) is weighted
 with the
 probability to find
 the
 fluctuation $h$ in the photon. $\langle
 T_A
 \rangle$ is the mean nuclear
 thickness function of
 the nucleus.
 The
 nuclear formfactor $F_A(q_L)$
 depends on the longitudinal momentum
 transfer
 in
 the diffractive dissociation
 $\gamma N \ss hN$.
 
 In what follows, we 
 try to
 justify our choice
 of the scaling variable and
 provide
 model-dependent
 estimates of $n(x,Q^2,A)$.
 \medskip

  2. {\bf A model for
 $n(x,Q^2,A)$}\\
At high energies the lifetime
of the hadronic fluctuations of the photon
may substantially exceed the nuclear radius,
$2\nu/Q^2\gg R_A$. In this case, the nuclear 
photoabsorption cross section can be represented
in the same form as in the hadron-nucleus interaction
 \cite{kl78,zkl,nz91},
 
 \beq
 \sigma^{\gamma^*A}_{tot}(x,Q^2)=
 2\int d^2b\left\langle 1-
 \left[1-
 \frac{\sigma(\rho,x)T(b)}
 {2A}\right]^A \right\rangle
\label{4}
\eeq

Here  $T(b)\approx
 \int_{-\infty}^{\infty}dz\rho_A(b,z)$ is the
 nuclear thickness function,
 where $\rho_A(b,z)$ is
 the nuclear density, which depends on the impact
 parameter $b$ and the longitudinal coordinate $z$.
The dipole cross section $\sigma(\rho,x)$ 
of the interaction with a nucleon of a $q\bar q$ pair
depends on its transverse separation $\rho$ 
and the energy related to $x=Q^2/2m_N\nu$. 
 The averaging over $\rho$ and $\alpha$, 
the fraction of the photon
 light-cone momentum carried by the quark,
is weighted with the photon
 wave function squared. 
One should be cautious
 with such a definition of
 the averaging \cite{nz91} in the case
of the photon, because its wave function, being 
different from the hadronic one, is not
 normalized to one. At 
small $\rho$ it has a form \cite{nz91} 
$|\Psi_{\gamma^*}^2(\rho,\alpha)|_T^2\ =
\newline 
6\alpha_{em}/(2\pi)^2\Sigma_f
 e_f^2 [1-2\alpha(1-\alpha)]\
 \epsilon^2K_1^2(\epsilon\rho)$, where
 $\epsilon^2=\alpha(1-\alpha)Q^2+m_q^2$.
The mean
 transverse separation of the $q\bar q$
 fluctuation is given by
 $\langle \rho^2 \rangle \propto 1/\epsilon^2$ and is
 of the order $1/Q^2$. However,
 at the endpoints of the
 kinematical region $\alpha$ or $1-\alpha\sim
 m_q^2/Q^2$, the $q\bar
 q$ fluctuations acquire
 a large transverse
 size, $\rho^2\sim 1/m_q^2$
 \cite{nz91,bk,fs1}.  For light quarks such a
 big size may substantially
 exceed the confinement
 radius, and one
 should put a cut off on the
 integration over $\alpha$.  This is equivalent
 to a replacement of the quark mass
 by the cut off $\lambda$.

 Expanding eq.
 (\ref{4}), we represent the nuclear
photoabsorption cross section 
in the form
 
 \beq
 \sigma^{\gamma^*A}_{tot}(x,Q^2)=
 A\ \sigma^{\gamma^*N}_{tot}(x,Q^2)
 \left
 [1-n(x,Q^2,A) + \ ...\right ]\ ,
\label{5}
\eeq
 where  $n(x,Q^2,A)=\langle T(b)\rangle
 \langle\sigma^2\rangle/4
\langle\sigma\rangle$ and
$\langle T(b)\rangle=(A-1)/A^2\int d^2bT^2(b)$.
 
 Note that
 expansion (\ref{5}) looks similar to
 that given by the
 Glauber
 approximation
 \cite{glauber}.  In fact we use the eikonal
 Glauber
 formalism for projectile states
 with definite transverse
 dimension
 $\rho$, since
 they are the eigenstates of interaction.  
However, as we conclude below,
 after averaging over the photon wave function
the first shadowing
 correction is of the order of $1/\lambda^2$, rather than
 $1/Q^2$ as
 expected in 
 the Glauber model. This comparison
 demonstrates that, in
 terms of multiple scattering
 theory, DIS on nuclei is
 dominated by Gribov's
 inelastic shadowing \cite{gribov}, while the
 Glauber eikonal
 contribution \cite{glauber}
 vanishes at high $Q^2$.
In such a case $n(x,Q^2,A)$ should be interpreted as
a measure of a number of gluons probed by the
$q \bar q$ fluctuation of the photon, rather than
a number of nucleons. This is justified
at small $\rho$ since the photoabsorption cross
section is proportional to $\rho^2$ and the gluon
distribution function \cite{barone,fs}. The latter
is not proportional to the nucleon density because 
of gluon fusion 
\cite{levin} - \cite{book}.
This effect is related to the inelastic corrections
corresponding to the excitation of heavy mass
intermediate states. On the other hand, if
$\rho$ is not small, the $q \bar q$ fluctuation
experiences additional shadowing interacting with
gluons. This results in a rather small unitarity
correction to the photoabsorption cross section on
a nucleon, but in a substantial correction on nuclei.
Taking into account the shadowing corrections coming
from large size fluctuations, we may say that such 
$q\bar q$ fluctuations also probe the number of gluons.

In order to evaluate the lowest order nuclear
 shadowing
 correction $n(x,Q^2,A)$
in eq. (\ref{5}), note that
 the
 denominator in eq. (\ref{1})
 is directly related to
the
 proton structure function,
 $\langle
 \sigma(x,\rho) \rangle
 = 4\pi^2\alpha_{em}\ F_2^p(x,Q^2)/Q^2$.
 We performed a fit to 
 available data on $F_2^p(x,Q^2)$
 from NMC
 \cite{nmc-p},
 H1 \cite{h1,h1-new} and ZEUS
 \cite{zeus,zeus-new}
 experiments with $x \leq 0.05$ 
 and $Q^2 \geq 0.5/ GeV^2$.
 We used a
 simple parameterization motivated by the
 double--leading--log approximation
 (DLLA) for QCD
 evolution equations
 \cite{book}
 (see also review
 \cite{review}),
 $F_2^p(x,Q^2)=f(Q^2)[a 
\exp(2\sqrt{L})/L +
 b\sqrt{x}]$,
 where the first term corresponds to the 
sea- quark (Pomeron),
 while the second
 term originates from the valence
- quark (Reggeons)
 contributions.
  $L=(4\pi/\beta_0) \ln(c/\alpha_s)
 \ln(d/x)$, where
 $\alpha_s=4\pi/\beta_0\ln(Q^2/\Lambda_{QCD}^2)$
 and $\Lambda_{QCD}=0.2\
 GeV$,
 $\beta_0=9$ for three active flavors.
 The factor
 $f(Q^2)=Q^2/(e+Q^2)$ guarantees
that the structure function
vanishes in the
 limit
 of real photoabsorption.
 This parameterization 
 fits the data very well 
 ($\chi^2/d.f. = 0.8)$ with
 parameters $a=0.036\pm 0.005\,
\ b=0.4\pm 0.08,\ c=0.59\pm 0.02,\
 d=0.31\pm 0.07,\ e=0.12\pm 0.08\ GeV^2$.
 
 Although DLLA
 does not provide a Regge form of
 the
 structure functions, an effective
 Regge
 parameterization may be a good
 approximation,
 $F_2^p(x,Q^2)
 \propto \exp[\Delta_{eff}(Q^2)\xi]$.
 Here
 $\xi=\ln(1/x)$ and
 $\Delta_{eff}(Q^2) =
 d\ln[F_2^p(x,Q^2)]/d\xi$
 correspond to the
 effective Pomeron intercept $\alpha_{eff} =1 +
 \Delta_{eff}$. Of course, 
 $\Delta_{eff}$ can be
 treated as $x$-independent
 only in a restricted
 interval of $x$.  The values of $\Delta_{eff}(Q^2,x)$
 corresponding to
 the results of our
 fit are depicted in fig.1 as function of $Q^2$
versus $x$. We see that $\Delta_{eff}(Q^2,x)$ 
is almost $x$-independent which justifies the Regge
parameterization as a good approximation
in the range of $x$ and $Q^2$ under consideration.
 
\begin{figure}[tbh]
\includegraphics{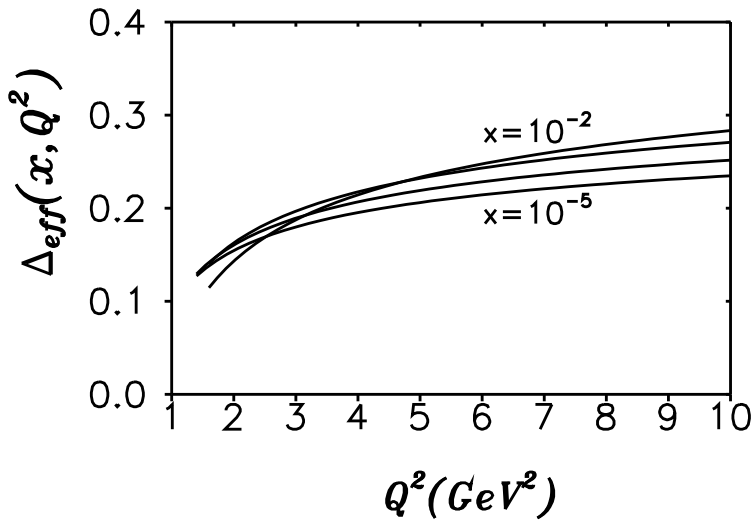}
\begin{center}
\vspace{7.5cm}
\parbox{13cm}
{\caption[Delta]
{The effective Pomeron intercept as function of
$Q^2$ versus\\ $x=10^{-2,3,4,5}$. The proton structure function
$F_2^p(x,Q^2)$ is fitted to available data as
is explained in the text.}}
%\label{event}}
\end{center}
\end{figure}

 Remarkably, the values of $\Delta_{eff}$ in
 fig. 1
 substantially exceed what is known from the
 energy dependence of
 the total
 cross sections of
 proton-proton interaction \cite{dklt,dl}, 
$\Delta_{eff} \approx 0.07 -
 0.08$. Our results
 show that this distinction
 remains substantial down to
 quite low $Q^2 \sim 2\ GeV^2$.
 Note that the observed $Q^2$-dependence 
of $\Delta_{eff}$ contradicts the Pomeron 
factorization.
This is not surprising since perturbative
QCD calculations \cite{bflk,nzz} show that
the Pomeron is a more complicated singularity,
a cut or a sequence of poles.

 The rising
 $Q^2$-dependence of
 $\Delta_{eff}(Q^2)$
means that the  $x$-dependence of the cross
 section $\sigma(\rho,x)$ is steeper at smaller
 $\rho$.
 Indeed, the larger $Q^2$ is, the smaller is
 $\langle \rho^2 \rangle
 \propto
 \ln(Q^2/\lambda^2)/Q^2$.
 
 Despite the smallness 
 of the mean
 transverse
 size of the photon fluctuations
 participating in
 DIS, of the order of $\sim 1/Q^2$,
 the shadowing terms in
 the
 expansion eq. (\ref{5}) are dominated by
 large
 transverse separations
 in $q\bar q$ fluctuations. 
This can be argued using 
 the relation,
 $\langle
 \rho^4 \rangle / \langle \rho^2
 \rangle =
 2.4 /\lambda^2
 \ln(Q^2/\lambda^2)$,
 which is obtained by 
the same averaging procedure 
as is defined in eq.  (\ref{4}). 
This relation is
 the manifestation of a salient feature of the
 hadronic fluctuations of
 the photon, namely, a huge
 dispersion of the transverse size
 distribution,
 $\langle \rho^4 \rangle \gg \langle \rho^2
 \rangle^2$.   Although we used the perturbative 
photon wave function, this conclusion has a 
rather general character. 
It is a result of the interplay of perturbative and
nonperturbative contributions in the deep-inelastic 
cross section. The former results from the small-size
fluctuations corresponding to 
"symmetric pairs", $\alpha \sim 1-\alpha$. They 
are presented with large probability in the 
photon wave function, but have a small,
$\sim 1/Q^2$ interaction cross section.
On the contrary, the highly asymmetric fluctuations
 with
 $\alpha$ or $(1-\alpha) \sim \lambda^2/Q^2$,
have a very small, $\sim 1/Q^2$ weight in the
photon wave function, but a large interaction
cross section,
typical for hadrons.
In the case of double scattering in the nucleus
the perturbative contribution turns out to be 
very much suppressed by the factor of $1/Q^4$, while
the nonperturbative part is suppressed only once 
by the weight factor of $1/Q^2$. 
Thus, the soft interaction 
dominates nuclear shadowing.

Actually, just this effect is responsible for
the  scaling  behavior of
 nuclear shadowing \cite{nz91}
and of unitarity corrections to the 
photoabsorption cross section 
on a nucleon\footnote{This
 conclusion is in variance with
 the statement in
\cite{kaidalov} that at
 high $Q^2$ the unitarity corrections
vanish as $\sim 1/Q^2$
 and one sees the single
 Pomeron
 exchange.}.  
 
 Once the interaction
 responsible for shadowing
 is essentially soft, its
 $x$-dependence is governed by the
 soft
 $ \Delta_{eff}(\lambda^2) \approx 0.1$,
 rather than
 the hard one.
 This is confirmed by the recent
 study of
 diffractive dissociation by the
 H1
 collaboration \cite{h1-dd}, which
 claimed
 $\Delta_P = 0.1 \pm 0.03 \pm 0.04$.
 
To proceed further with the 
calculation of $n(x,Q^2,A)$,
note that in eqs.
 (\ref{4}), (\ref{5}) 
we temporarily used an
  assumption 
 that
 the photon energy
 $\nu=Q^2/2m_Nx$ in the nuclear rest frame is
 sufficiently
 high to make the lifetime of the
 photon fluctuation long
 compared with the
 nuclear
 size, so that it propagates through the whole
 nucleus
 with a frozen
 intrinsic separation $\rho$.
 However, most of the data
 available are in the
 transition  region of $x$, 
where the lifetime, usually
 called 
 coherence time,
 is comparable with the nuclear
 radius.  The finite  coherence time
 can be taken into account by
 introducing a phase shift between $q\bar q$
 wave
 packets produced at
 different longitudinal
 coordinates, in the same
 way as for
inelastic corrections \cite{kk}, or in the
 vector dominance model \cite{bauer}.
The mean nuclear
 thickness function of
 eq.  (\ref{5}) should be replaced
by an effective one,
 
 \beq
 \langle\widetilde{T}(b)\rangle={A-1\over A^2}\int d^2b
 \left[\int_{-\infty}^{\infty}dz\
 \rho_A(b,z)\ e^{iqz}\right]^2
 \approx\langle
 T(b)\rangle\ F_A^2(q_L)\ .
\label{8}
\eeq
Here $F_A(q_L)=\exp(-q_L^2R_A^2/6)$ is the
 nuclear longitudinal
 formfactor and $R_A$ is the mean
 nuclear radius.  For the sake of simplicity
 we
 use
 the Gaussian form for the nuclear density which is
 quite precise
 for the 
 calculation of $F_A(q_L)$.  Calculating
$\langle T \rangle$ we use the realistic 
 parameterization of nuclear density \cite{jager}.

 The decrease of the effective
 nuclear thickness
 function $\langle\widetilde{T}(b)\rangle$ at large
 $q_L$
 can be interpreted as a result of shortness
 of the hadronic
 fluctuation path
 in the nucleus, if
 we are in the nuclear rest frame, or as
 an
 incomplete
 overlap of the gluon clouds of the
 nucleons which have the
 same impact
 parameter in
 the infinite momentum frame of the nucleus
 \cite{kancheli,nz-old}.
 
 In order to calculate the longitudinal momentum
 transfer in DIS,
 $q_L=(Q^2+M^2)/2\nu$, one needs
 to know the effective
 mass of the produced
 $q\bar
 q$ wave packet.  However, a $q\bar q$ state
 with
 definite separation
 $\rho$ does not have a
 definite mass. This is a
 typical problem for
 those
 who work in the eigenstate basis of
 interaction.  We evaluate $q_L\approx
 2xm_N$
 assuming $M^2\sim Q^2$.  Thus,
 the parameter
 which controls the
 value of
 $\langle\widetilde{T}(b)\rangle$
is $x$.
 
 Now we are in a position to
 estimate
 the nuclear
 shadowing correction
per nucleon $n(x,Q^2,A)$
in eq. (\ref{5}), which reads
 
 \beq
 n(x,Q^2,A)={1\over 4} 
 \frac{N}{F_2^p(x,Q^2)}
 \langle
 T(b)\rangle F^2_A(q_L) \left({1\over
 x}\right)^{2\Delta_{eff}(\lambda^2)}\ .
\label{10}
\eeq
The scaling variable $n(x,Q^2,A)$
 can be interpreted as a measure of the amount of
 those gluons which take part
 in
 the interaction with the
 $q\bar q$ fluctuation during its lifetime.
 Eq. (\ref{10}) is a model-dependent realization 
 of the general expression (\ref{1}). 
 
 \medskip
 
 3. {\bf Comparison with the data}\\
 The  variable
 $n(x,Q^2,A)$ has
 been calculated
 from eq.  (\ref{10}) using
 the
 values
 of $x$, $Q^2$ and $A$ corresponding to
 data
 from the NMC
 experiment
 \cite{nmc1,nmc2} as well as the
 results of
 our fit to
 $F_2^p(x,Q^2)$.  The expected
 scaling dependence
 of the nuclear shadowing
 on
 $n(x,Q^2,A)$ is not affected
 by the overall
 normalization factor $N$.
However, in order to have 
a correct slope of $n$-dependence 
corresponding to eq. (\ref{5})
we choose $N=3\ GeV^{-2}$.
 The data on
 the ratio of the nuclear and
 nucleonic
 photoabsorption cross sections
 $R_{A/N}(x,Q^2)$ is
 plotted
 against 
 $n(x,Q^2,A)$ in
 Fig.2.  They demonstrate a good
 scaling in
 $n(x,Q^2,A)$ 
 within a few percent accuracy.

\begin{figure}[tbh]
\includegraphics{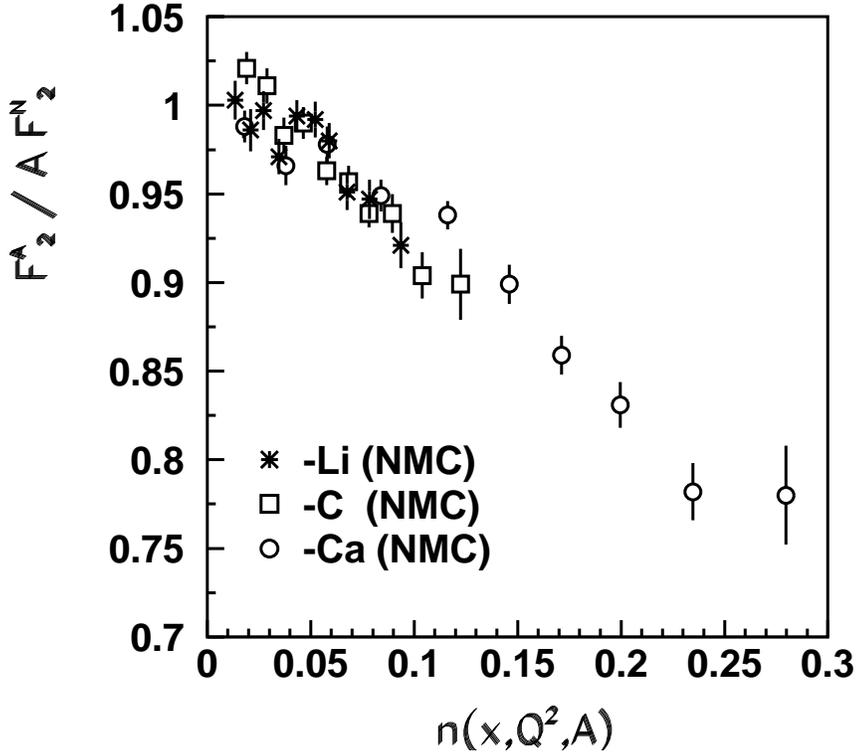}
\begin{center}
\vspace{10cm}
\parbox{13cm}
{\caption[Delta]
{Data on nuclear shadowing at small $x$ from the
 NMC
 \cite{nmc1,nmc2} experiments versus the scaling
 variable
 $n(x,Q^2,A)$ as defined in eq.  (\ref{10}).  The
slope of the  straight line
 corresponds to the normalization factor in 
eq. (\ref{10}) $N=3\ GeV^{-2}$.}}
%\label{event}}
\end{center}
\end{figure}

 We should comment more on the procedure
 of the calculation
 of $n(x,Q^2,A)$:
 
 {\sl(i)} Our considerations are
 valid only for small
 $x$, so we limit the
 $x$-region to $x<0.07$.  At
 larger
 $x$, the nuclear
 structure functions show
 a small enhancement of a few percent
relative to
 the proton one, which
 results in
 $R_{A/N}(x,Q^2)>1$ for $n(x,Q^2,A) \ss 0$.  
 A plausible assumption is that about
the same 
 antishadowing correction extends
 down to 
 smaller $x$, where it is compensated by stronger 
 shadowing
 effects. Such a behavior, for instance,
is expected 
 in the model of swelling bound
 nucleons \cite{close,barone}.
 The antishadowing effect may have some $A$
 dependence, what
 would cause a small,
 a few percent relative shift of the
 data
 in fig. 2 corresponding to different nuclei, but
will not change 
 the slope
 of $n$-dependence.
 Since the physics of
 antishadowing is beyond the scope of
 our present consideration, and the
 effect is numerically
 very small, we do not try to incorporate with it,
 but just have renormalized the solid line 
 $R_{A/N}(n)=1-n$ in fig. 2
 by 3\% up to make the
 comparison easier.
 
 {\sl(ii)} The data points
 \cite{nmc2} for
 $Q^2<0.5\ GeV^2$ were
 excluded from the analysis
 because they are in the
 realm of the vector
 dominance model, rather than DIS.
 They should
 correspond to the same nuclear
 shadowing
 as is experienced by the
 $\rho$-meson. This
 is the reason for the
 saturation of
 nuclear shadowing at small $x$,
 claimed
 in \cite{e665,nmc2}.
 
 {\sl (iii)} A further important observation is that
 $R_{A/N}(x,Q^2)$ depends to a good accuracy
 linearly upon
 $n(x,Q^2,A)$ at
 least for $n<0.2$.  
 The higher order
 terms in
 the expansion in eq.
 (\ref{5}) are expected
 to violate the linearity in $n$. 
 This implies that
 those terms are small.
 A model-dependent evaluation of the next
 shadowing
 correction shows that it is small indeed.
 
 {\sl (iv)} 
 There are two
 contributions to the shadowing in
 DIS \cite{levin,mueller,qiu}, one
 comes from the suppression of
 the gluon
 density as a consequence
 of gluon
 fusion $gg\ss
 g$, which
 corresponds to the triple Pomeron graph
 in the
 framework of standard Regge
 phenomenology.
 Another contribution to
 the
 shadowing comes from
 the
 Glauber-like rescattering of the $q\bar q$
 fluctuation off gluons.  The
 latter mechanism, which was
 mostly under
 consideration
 above, can also
 be
 viewed upon as a parton fusion, but as a
 fusion of
 gluons into a
 $q\bar q$ pair, $gg\ss q\bar q$. In
 the Regge-model
 this process corresponds
 to the Pomeron-Pomeron-Reggeon graph.
 Both
 mechanisms lead to the same form of
the variable $n(x,Q^2,A)$ in eq. (\ref{10}).
However,
the formfactor $F_A(q_L)$ has a different
form for the
 triple-Pomeron mechanism
 due
 to the contribution of heavier hadronic
 fluctuations of
 the photon,
 $F_A(q_L)\propto Ei(-x^2m_N^2R_A^2/6)$.
 Here $Ei$ is the integral exponential
 function. 
 We checked that
 in the $x$ and $A$ domain investigated, the
 admixture of the
 triple-Pomeron mechanism does not
  affect the
 $n(x,Q^2,A)$-scaling within
the error bars of the data available.
It may, however, cause a deviation 
from the scaling for heavy nuclei.
We hope that  forthcoming high-statistic
 data on heavy nuclei 
from the NMC Collaboration
 may help to disentangle
 these two mechanisms of shadowing.
This is important if one wants to predict 
the unitarity corrections
to the proton structure function at small $x$
or the photon diffractive dissociation
cross section, because 
the admixture of the triple-Pomeron  
affects the normalization constant $N$ in
eq. (\ref{10}). 
\medskip
 
 {\bf 4. Summary}\\ 
Starting from the QCD dynamics of
deep-inelastic scattering at small $x$ 
we have found a new
 variable
 $n(x,Q^2,A)$ which scales all available data
 on nuclear
 shadowing in
 DIS  at small-$x$.  This
 variable 
 measures the number
 of gluons
 which a hadronic fluctuation of the virtual
 photon
 interacts
 with during its lifetime.

An important observation is also that 
shadowing corrections
at small $x$ and large $Q^2$ grow
less steeply than $[F_2^p(x,Q^2)]^2$.
This is because 
nuclear shadowing is a subject
to soft rather than hard physics.

The observed scaling of nuclear shadowing
as function of $n(x,Q^2,A)$ 
supports our assumptions on 
the dynamics of nuclear shadowing.

 {\bf
 Acknowledgement:} We would like to thank
 J.~H\"ufner, P.~Landshoff and
 E.~Predazzi for
 useful discussions, and C.~Schr\"oder, who
read the manuscript and made usefull improving suggestions.
 B.K. thanks the MPI f\"ur
 Kernphysik,
 Heidelberg,
 for financial support.

\end{document}